\pdfoutput=1

\documentclass[conference, 10pt]{IEEEtran}
\IEEEoverridecommandlockouts
\def\BibTeX{{\rm B\kern-.05em{\sc i\kern-.025em b}\kern-.08em
    T\kern-.1667em\lower.7ex\hbox{E}\kern-.125emX}}

\usepackage[bottom=0.82in, top=0.70in, left=0.625in, right=0.64in]{geometry}

\ifCLASSINFOpdf
   \usepackage[pdftex]{graphicx}
   \graphicspath{{./}{./img/}}
   \DeclareGraphicsExtensions{.pdf,.jpeg,.jpg,.png,.eps,.ps}
\else
  \usepackage[dvips]{graphicx}
  \graphicspath{{./}{./img/}{./eps/}}
  \DeclareGraphicsExtensions{.pdf,.jpeg,.jpg,.png,.eps,.ps}
\fi
\usepackage{cite}
\usepackage[cmex10]{amsmath}
\usepackage{algorithmic}
\usepackage{array}
\usepackage{mdwmath}
\usepackage{mdwtab}
\usepackage{eqparbox}
\usepackage[font=footnotesize]{subfig}
\usepackage{stfloats}
\usepackage{url}

\usepackage{textcomp}
\usepackage{xcolor}
\def\BibTeX{{\rm B\kern-.05em{\sc i\kern-.025em b}\kern-.08em
    T\kern-.1667em\lower.7ex\hbox{E}\kern-.125emX}}
    
\usepackage[lined, ruled, linesnumbered]{algorithm2e}
\usepackage[bookmarks=false]{hyperref}
\usepackage{latexsym}
\usepackage{amsfonts,amssymb,amsthm}

\usepackage{enumitem}
\usepackage{pbox}
\usepackage{mathtools}
\usepackage{float}
\usepackage{pgffor}
\usepackage{bm}		% \bm command in math mode
\usepackage{tabularx}	% tabularx environment
\usepackage{framed}	
\usepackage{cleveref}	% \subref command
\usepackage{pifont}		% \cmark and \xmark defined in cmd.tex
\usepackage{makecell}
\usepackage{graphicx}
\usepackage{diagbox}
\usepackage{float} 
\usepackage{booktabs}
\usepackage{multirow}
\usepackage{setspace}
\usepackage{amsmath}
\usepackage[T1]{fontenc}

\IEEEoverridecommandlockouts

\IEEEaftertitletext{\vspace{0.3mm}}

\newtheorem{definition}{\bf Definition}	%[]
\newtheorem{theorem}{\bf Theorem}		%[]
\newtheorem{property}{\bf Property}	%[]
\newcounter{appdx}
\definecolor{dgreen}{rgb}{0,0.655,0.149}

\newcommand{\redc}[1]{{\color{red}}}

\SetKw{Continue}{continue}
\SetKw{Break}{break}

\makeatletter
\newcommand{\nosemic}{\renewcommand{\@endalgocfline}{\relax}}% Drop semi-colon ;
\newcommand{\dosemic}{\renewcommand{\@endalgocfline}{\algocf@endline}}% Reinstate semi-colon ;
\let\oldnl\nl% Store \nl in \oldnl
\newcommand{\nonl}{\renewcommand{\nl}{\let\nl\oldnl}}% Remove line number for one line
\makeatother

\SetKwInOut{Init}{Initialize}
\SetKwInOut{OptIn}{Optional input}

\begin{document}

\title{Moving Edge for On-Demand Edge Computing: An
Uncertainty-aware Approach\\

}
\author{Fangtong Zhou and Ruozhou Yu %, Guoliang Xue
\thanks{Zhou and Yu  (\{fzhou, ryu5\}@ncsu.edu) are with the Department of Computer Science at the NC State University, Raleigh, NC 27606, USA.
%
%
% Xue (xue@asu.edu) is with the School of Computing and Augmented Intelligence at the Arizona State University, Tempe, AZ 85287, USA.
%
%
This research was supported in part by NSF Grants 2045539 and 2433966.
The information reported here does not reflect the position or the policy of the funding agency.
}
}

\maketitle

\begin{abstract}
We study an edge demand response problem where, based on historical edge workload demands, an edge provider needs to dispatch \emph{moving computing units}, \emph{e.g.} truck-carried modular data centers, in response to emerging hotspots within service area.
The goal of edge provider is to maximize the expected revenue brought by serving congested users with satisfactory performance, while minimizing the costs of moving units and the potential service-level agreement violation penalty for interrupted services.
The challenge is to make robust predictions for future demands, as well as optimized moving unit dispatching decisions.
We propose a learning-based, uncertain-aware moving unit scheduling framework, \textbf{URANUS}, to address this problem.
Our framework novelly combines Bayesian deep learning and distributionally robust approximation to make predictions that are robust to data, model and distributional uncertainties in deep learning-based prediction models.
Based on the robust prediction outputs, we further propose an efficient planning algorithm to optimize moving unit scheduling in an online manner.
Simulation experiments show that URANUS can significantly improve robustness in decision making, and achieve superior performance compared to state-of-the-art reinforcement learning, uncertainty-agnostic learning-based methods, and other baselines.
\end{abstract}
\begin{IEEEkeywords}
Edge Computing, 
Edge Demand Response,
Resource Allocation, 
Bayesian Deep Learning, Distributional Robustness
\end{IEEEkeywords}

\section{Introduction}
\label{sec:intro}

\noindent
Edge computing (EC) brings computation and data storage close to data sources~\cite{ec_cha}, with the promise to reduce network congestion and latency brought by the massive number of intelligent mobile devices with excessive data processing demands~\cite{MOBIHOC20}.
However, as EC has limited resources, it is challenging to accommodate frequently changing edge demands.
Specifically, fixed, static deployment of edge resources may lead to alternating under-utilization and over-utilization of edge resources.
Under-utilization increases edge provider (EP)'s cost without increasing revenue, while over-utilization affects user experience which also degrades revenue.
Further, if the EC service becomes unavailable due to congestion and resource over-utilization, the EP may incur high penalty for violating \emph{service-level agreements (SLAs)}.

Responding to fluctuating edge demands is key to ensuring system performance and user experience~\cite{resource_allocation}.
Many solutions rely on estimating future edge demands before decision making, based on either assumptions such as \emph{identically and independently distributed (IID)} demands over time~\cite{MOBIHOC20}, or predictive models such as \emph{deep neural networks (DNNs)} to predict future demands~\cite{baochun_infocom21, zhang_infocom21, edp_bicycle, edp_cnn}.
Nevertheless, as edge demands are intrinsically uncertain and dynamic, methods based on the IID assumption are not robust against demand changes over time.
Meanwhile, predictive models such as DNNs could lead to overconfident predictions and bring model uncertainty as evidenced in existing works~\cite{variational}.
Eventually, both lead to inaccurate estimations, and cause SLA violations when resources are insufficiently provisioned for the demands.

This paper makes two contributions to tackle the above issues.
On the architecture side, we propose the concept of the \textbf{Moving Edge} (MOE).
Instead of solely relying on static EC resources~\cite{resource_allocation, INFOCOM22_2}, such as edge servers with base stations, the EP deploys additional \emph{moving computing units} (MUs) to provide \emph{on-demand response} to edge demand surges.
On the algorithm side, we further study how to schedule MU resources across locations to maximize the EP's profit.
In face of potential SLA violations and penalty, we design a general framework for \emph{uncertainty-aware} demand prediction and MU scheduling.
Our framework, called \textbf{URANUS}, novelly combines 1) Bayesian deep learning for uncertainty-aware edge demand prediction, 2) distributionally robust approximation for robust revenue/penalty estimation, and 3) finite-horizon multi-MU planning for on-demand scheduling.
To evaluate our framework, we conducted simulation experiments based on real-world data.
Results show that by explicitly considering uncertainty and robustness in online decision making, our framework can notably outperform a number of baselines, including: end-to-end reinforcement learning, uncertainty-agnostic decision making, and greedy heuristics.
Our contributions are as follows: 
\begin{enumerate}[leftmargin=1.4em]
    \item 
    We design URANUS, an uncertainty-aware MU scheduling algorithm, which takes the historical edge demands as input and makes online, uncertainty-aware scheduling decisions.
    \item 
    We develop a Bayesian neural network-based time series prediction model to predict the future demands with uncertainty quantification.
    \item 
    To address distributional uncertainty in Bayesian variational approximations, we develop a distributionally robust conservative estimation, solved with semi-definite programming.
    \item 
    We then propose a finite-horizon single-agent optimal planning algorithm given prediction and estimation outputs, and extend it to a multi-agent algorithm.
    \item 
    We conducted extensive evaluations of various algorithms using the real-world data from Telecom Italia, which demonstrate the superior performance of our framework.
\end{enumerate}

This paper is organized as follows. 
Sec.~\ref{sec:rw} talks about related work. 
Sec.~\ref{sec:model} describes the edge demand model and problem formulation. 
Sec.~\ref{sec:overview} gives an overview of the URANUS framework.
Sec.~\ref{sec:bnn} tackles prediction of uncertain demands. 
Sec.~\ref{sec:risk} describes the distributionally robust approximation.
Sec.~\ref{sec:dp} presents the finite-horizon planning algorithm.
Sec.~\ref{sec:eval} shows the evaluation results. 
Sec.~\ref{sec:conclusions} concludes this paper.

\section{Related Work}
\label{sec:rw}

\noindent
\textbf{Edge resource allocation and demand
response.}
\noindent
Edge resource management has received extensive efforts in existing research, such as recently in~\cite{MOBIHOC20, INFOCOM21_0, INFOCOM22_2, ra_1, ra_2, ra_3, Li2022b, Wang2022c, Chen2022c} and many more, with a wide variety of algorithmic, optimization and learning-based methods proposed.
However, a few major challenges still exist in the literature.
First, many assume full knowledge of demands as input to resource optimization formulations~\cite{ra_3, INFOCOM22_2, Li2022b}, or unknown but static demands which can be easily learned by a bandit or learning algorithm~\cite{INFOCOM21_0, ra_1, ra_2}.
Online algorithms~\cite{Chen2022c, Wang2022c} assume no knowledge of demands, leading to overly conservative solutions.
In reality, edge demands are highly fluctuating and dynamic.
Second, prediction-based methods assume the prediction models are accurate, while in reality prediction models such as DNNs bear notable data and model uncertainties~\cite{INFOCOM21_0}.
Third, learning-based decision making such as bandit algorithms~\cite{ra_1, ra_2} or deep reinforcement learning (DRL)~\cite{INFOCOM21_0} require long time and enormous data to converge.
Many existing works use fixed EC resources, which can be inefficient and insufficient for fluctuating edge demands.

\noindent
\textbf{Moving edge.}
\noindent
The inefficacy of fixed edge resource deployment has led to recent studies on movable EC resources.
An example is unmanned aerial vehicle (UAV)-assisted mobile EC~\cite{MOBIHOC22_0, INFOCOM22_0}, where computing power is offered by flying UAVs to performance-stringent applications.
While UAVs are more flexible in deployment, they have very limited energy budget and computing power, and can barely provide non-interrupted high-performance computing services for an extended period.
Instead, modular data centers (MDCs)~\cite{MDC} carried on large moving vehicles may be able to offer much more performant and stable services with support from stable ground power source and Internet access.
Compared to the UAVs with average flight time of 30-40 mins and charging time of 60-90 mins~\cite{drone_charge}, large vehicles carrying MDCs can provide days of constant driving until refueling.
MDC purchasing or renting services are widely provided by large companies such as Microsoft~\cite{Azure_MDC}, Dell~\cite{Dell_EMC}, Google~\cite{Google_PMDC} and HP~\cite{HP_POD}.
Additionally, services for data and computing on-the-wheels are being provided by companies like Amazon~\cite{amazon_data_wheeler}, IBM~\cite{ibm_data_transport} and BROADCOM~\cite{broadcom_data_transport}.
These existing services strongly motivate future portable computing centers hosted on ground vehicles for rapid demand response.

\noindent
\textbf{Edge demand prediction.}
\noindent
Existing works on edge demand predictions mostly focus on canonical DNNs.
In~\cite{baochun_infocom21, zhang_infocom21}, long short-term memory (LSTM) networks are used to predict wireless traffic demand to realize intelligent network operations.
A hybrid DNN framework with auto-encoder and LSTM is proposed in~\cite{wang2017spatiotemporal} to capture the spatial-temporal dependency in wireless traffic across cells.
In~\cite{nie2017network}, a deep belief network-based prediction method for wireless networks was proposed.
Our work differentiates from the above via a Bayesian approach to explicitly model and tackle uncertainties~\cite{bnn_survey}, and distributionally robust approximation to further improve robustness of decision making based on the predictions.

\noindent
\textbf{Robust optimization in edge computing.}
\noindent
Robust and distributionally robust optimization methods have been applied in EC scenarios, such as edge offloading~\cite{qu2020robust, chen2021energy, tra_off_cvar}, resource provisioning~\cite{chen2020data,  MOBIHOC20, Zhao2021d}, etc.
Robust optimization assumes no prior knowledge and is typically too conservative for realistic settings~\cite{liu2019transactive}.
Distributionally robust optimization assumes prior knowledge such as an empirical distribution~\cite{chen2021energy}.
The novel integration of machine learning (ML) models and (distributionally) robust optimization has only been recently studied in the ML community under very restricted scenarios such as computer vision and robotics~\cite{rahimian2019distributionally}, and has not been applied in edge resource management.

\noindent
\textbf{Decision making in dynamic environments.}
\noindent
Decision making in dynamic environments has been traditionally tackled via two kinds of approaches: model-based and model-free.
A prevalent example of model-based decision making is Model Predictive Control (MPC)~\cite{mpc}, which employs a predictive model to predict the future environment and response of a control system, and implements an optimal control by solving an optimization problem over a finite time horizon.
MPC is widely used in various industrial control systems such as industrial robotics~\cite{mpc_rob} and autonomous driving~\cite{mpc_drive}, where robustness and stability are strongly preferred.
Model-free approaches such as reinforcement learning (RL)~\cite{drl} do not require a predictive model, but instead seek to directly learn the optimal policy from extensive training data and/or interactions with the environment.
While RL approaches can learn any optimal policy with ample data, it suffers from degenerate performance with limited data, or instability when the environment changes~\cite{comp_mpc_drl}.
The optimal choice between model-based and model-free control highly depends on the problem instance and requirement.

\section{System Model}
\label{sec:model}

\subsection{Edge Demand Model}
\label{sec:demand_model}

\noindent
Let $\mathcal{A}$ be the geographical area covered by the EC service of an EP, divided into service cells $\mathcal{A} \!=\! \{ a_1, a_2, \dots, a_n \}$.
Each cell represents a subarea where users are served by the same fixed EC node.
A cell is equipped with a set of fixed resources provided by the EC node (\emph{e.g.}, the base station), such as CPUs, GPUs, RAM and bandwidth, denoted by $\mathbf{r}_a \!=\! (r_a^1, r_a^2, \dots, r_a^K) \in \mathbb{R}_*^K$ where $K$ is the number of resource types and $\mathbb{R}_*$ is the set of non-negative real numbers.

Let $\mathcal{T} \!=\! \{ 1, 2, \dots \}$ denote system time.
Let $d_{a}(t) \in \mathbb{R}_*$ be the \emph{EC workload (demand)} in cell $a \in \mathcal{A}$ at time $t \in \mathcal{T}$, coming from users of heterogeneous applications.
Serving the demand requires a minimum amount of computing and network resources, denoted by a function 
$\mathbf{f}_a: \mathbb{R}_* \rightarrow \mathbb{R}_*^K$, where $\mathbf{f}_a(d) \!=\! ( f^1_a(d), f^2_a(d), \dots, f^K_a(d) )$ denotes the minimum resources to serve $d \in \mathbb{R}_*$ units of demand.
We assume the minimum resources are constant or linear in the demand, \emph{i.e.}, $f_a^k(d) \!=\! \phi_a^k \cdot d + \varphi_a^k$ for $k = 1, 2, \dots, K$ where $\phi_a^k, \varphi_a^k \in \mathbb{R}_*$.
We also assume the fixed edge resource $r^k_a$ satisfies $r^k_a \ge \varphi^k_a$; or else the fixed EC node would not even serve any demand.

In reality, EC workloads are subject to frequent \emph{temporal-spatial fluctuations}~\cite{MOBIHOC20}.
Users' mobility and application usage patterns change based on work/off hours, time, and location. Additionally, workloads are influenced by events like holidays and accidents.
The effects are two-fold:
1) the demands are dynamic over time and location following some unknown trend;
2) the demands are uncertain for a specific time and location.
Formally, we model the demands $\mathbf{d}(t) = \{ d_a(t) \,|\, a \in \mathcal{A} \}$ as a stochastic process where $\mathbf{d}(t)$ at each $t$ follows some (unknown) distribution.
Hence $\mathbf{f}(t) = \{ \mathbf{f}_a(d_a(t)) \,|\, a \in \mathcal{A} \}$ is also a stochastic process that depends on the demands $\mathbf{d}(t)$.

\subsection{Moving Edge for On-Demand Computing}
\begin{figure}[tb]
\centering
{\includegraphics[width=0.9\linewidth, trim=230 230 270 170, clip]{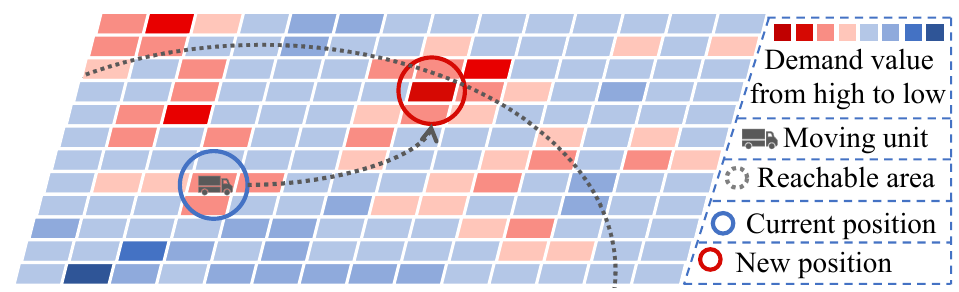}
}
\caption{Dispatching MUs based on demands}
\label{fig:map}
\end{figure}

\noindent
A static edge resource deployment can result in either resource wastage during demand troughs or over-utilization during demand surges, leading to degraded performance.
To solve resource under- and over-utilization, we propose to serve edge workloads \emph{on-demand} via {the MOE}.
The MOE consists of EC hubs deployed on moving vehicles (called MUs) that can be easily dispatched to various locations for quick response to demand surges.
The idea is inspired by recent developments in modular and mobile computing hubs, such as Azure Modular Data Center~\cite{azure}.
Unlike UAVs, MUs require supporting ground facilities for power, high-speed Internet connection, etc., but offer more powerful and stable computing than UAV-mounted nodes.
Fig.~\ref{fig:map} shows that an MU is dispatched to the place with the highest edge demand within reach.

The core of MOE lies in flexible, on-demand MU resource management in response to fluctuating edge demands.
This requires predicting edge demand surges across locations, and making MU scheduling decisions to serve geo-distributed hotspots.
In what follows, we formulate the MU scheduling problem.

\subsection{Problem Formulation}
\noindent
Consider an MOE system with $m$ homogeneous MUs.
Each MU can provide a certain amount of resources to local users, denoted by $\mathbf{g} = (g^1, g^2, \dots, g^K)$.
As a moving resource, an MU has three working modes: \emph{operational}, \emph{in transit} and \emph{idle}.
\textbf{Operational} (denoted by ``$\mathsf{opt}$'') means the MU is deployed at a ground facility in some cell $a \in \mathcal{A}$ and serving near-by users.
\textbf{In transit} (denoted by ``$\mathsf{tra}$'') means the MU is \emph{en route} to another location for re-deployment, during which it will not be able to provide EC service due to lack of a stable power source, Internet access or other required ground resources.
The transit time between two cells $a, b \in \mathcal{A}$ is $T_{ab}$.
\textbf{Idle} (denoted by ``$\mathsf{idl}$'') means the MU is deployed in a cell $a$ but is not operational to serve local users, in order to minimize the operational cost when the demand is low.
In the following, we formulate the operation of MUs as a {Markov Decision Process (MDP)}:
\begin{definition}[MU Scheduling]
\label{def:mu}
The MU scheduling problem in MOE-based on-demand EC is defined with the following:
\begin{itemize}[leftmargin=1em]
    \item \textbf{Agents:} The system has $m$ homogeneous MUs as agents.
    \item \textbf{States:} A state of the system includes two parts: the environment state $\mathcal{S}_{\mathsf{env}}$, and the system state $\mathcal{S}_{\mathsf{sys}}$.
    $\mathcal{S}_{\mathsf{env}}$ includes the current time $t$, and the historical edge demands $\{ \mathbf{d}(\tau) \,|\, \tau \le t \}$ in all cells by time $t$. 
    We assume $\mathcal{S}_{\mathsf{env}}$ (the demands) evolves over time independent of MU scheduling. 
    $\mathcal{S}_{\mathsf{sys}}$ includes modes and locations of all MUs, depending on scheduling decisions in the previous time steps.
    \item \textbf{Actions:} For each agent $i$, its action for time $t$ includes a mode $x_i(t) \in \{ \mathsf{opt}, \mathsf{tra}, \mathsf{idl} \}$.
    Specifically, an MU in modes opt or idl in the previous time step will stay in the same location and be available to schedule.
    An MU in transit to a new destination is not schedulable (cannot take any action) before it arrives at the destination, during which we must have $x_i(t) = \perp$ denoting no action.
    If $x_i(t) = \mathsf{tra}$, then additionally a destination $\delta_i(t) \in \mathcal{A}$ is designated; otherwise we have $\delta_i(t) = \perp$ meaning no destination. 
    Let $\mathbf{X} = \{ \mathbf{X}(t) \,|\, t\in\mathcal{T}\} = \{\{x_i(t), \delta_i(t) \,|\, i \} \,|\, t\in\mathcal{T}\}$ denote the actions (modes and destinations) decided for each MU $i$ at each time $t$, and let $\{ z_{\mathsf{opt}}(t), z_{\mathsf{tra}}(t), z_{\mathsf{idl}}(t) \}$ be the number of MUs in the three modes respectively based on $\mathbf{X}(t)$.
    \item \textbf{Rewards:} At each $t$, the EP earns profit which depends on the state and actions. We formulate it in the next subsection.
\end{itemize}
\end{definition}

\subsection{Edge Provider's Profit}

\noindent
Normally, an EP's profit is decided by two items: its net \emph{revenue} for serving users, and its \emph{cost} for providing the service.
An EP may also incur a \emph{penalty}, if it violates an SLA.

\textbf{System revenue (utility).}
The EP earns revenue from serving users with satisfactory performance, and bears revenue degradation when performance degrades.
For example, in online retail, every $100$ms of latency could reduce sales by $1\%$~\cite{amazon}.
For cell $a$, define system revenue at time $t$ as $\overline U_a(t) \!=\! \widehat U_a(t) + U_a(t)$, where $\widehat U_a(t)$ is the \emph{base revenue} for serving all users at time $t$, and $U_a(t)\!<\! 0$ is the \emph{(negative) utility} due to increased latency.
In practice, $\widehat U_a(t)$ is commonly a constant based on the service contract, independent of MU scheduling and omitted here.
$U_a(t)$ is monotonically non-decreasing with decreasing demand and/or increasing resources.
Let $\rho_a^k(t) = r_a^k + g^k \cdot z_a(t) - f_a^k(d_a(t))$ be the \emph{residual resource} of type $k$, which is the total fixed and MU resources in $a$ at time $t$, minus the minimum resource requirements $f_a^k(d_a(t))$.
Here $z_a(t)$ is the number of MUs being operational in cell $a$ at time $t$.
An example utility function used in  evaluation is:

\begin{equation}
\label{eq: utility} 
    U_a(t) \!=\! 
    \left \{
    \begin{array}{ll}
    \! -\min\left\{ \max_{k} \left\{ \frac{u_a^{k}}{  \rho_a^k(t) } \right\}, -U_{\sf base} \right\}, &\!\!\!\!\!\!
    \text{ if }\forall k, \rho_a^k(t) > 0 \\
    \! U_{\sf base}, &\!\!\!\!\!\! \text{ otherwise} 
    \end{array}
    \right.
\end{equation}

where $u_a^{k} \in \mathbb{R}_*$ is a constant value, and $-U_{\sf base}$ is an upper bound of the latency-incurred revenue loss, that is, the base revenue, which makes sure the EP is budget balanced.
Eq.~\eqref{eq: utility} draws intuition from viewing each resource as an independent queueing system, whose queueing latency is proportional to the reciprocal of $\rho_a^k(t)$, weighted by utility constant $u_a^{k}$.

\textbf{System cost.} 
Neglecting the costs for operating the fixed resources which are also constant to MU scheduling, we define $c_{\sf opt}$, $c_{\sf tra}$ and $c_{\sf idl}$ as the costs for each MU when working in the three modes, respectively.
The total cost at time $t$ is:
\begin{equation}
 \label{eq: cost}
    C(t) =  c_{\sf tra} z_{\mathsf{tra}}(t) + c_{\sf opt}  z_{\mathsf{opt}}(t) + c_{\sf idl}  z_{\mathsf{idl}}(t).
\end{equation}

\textbf{System penalty.} 
System penalty comes from some users not being served due to lack of resources.
Ideally if $\rho^k_a(t) \ge 0$ for $\forall a, k$, then all users will receive at least the basic service and no penalty is incurred.
But when too many users arrive and $\rho^k_a(t) < 0$ for some $a, k$, 
some users may have to lose access, or the service will not be able to be fully started.
With uncertain $d_a(t)$, it is hard to enforce a strict constraint on $\rho^k_a(t)$, and real-world SLAs are commonly defined with probabilistic terms, such as the service being ``$95\%$'' available to avoid penalty.
The SLA requirements can be expressed with a \emph{chance constraint}:
\begin{align}
\label{eq:pen_chance}
    \Pr \left[ \bigcap_{k = 1}^K \left( \rho^k_a(t) \ge 0 \right) \right] \ge 1 - \epsilon,
    \;
    \forall a, t,
\end{align}
where $\epsilon$ is a pre-negotiated tolerance threshold for the SLA.

When the SLA is violated, the {{{EP}}} will incur a penalty commonly \emph{proportional to} the severity of the violation~\cite{sla_proper}, \emph{i.e.}, how much demand loses access to the edge service in our case.
This value, defined as the \emph{expected excess demand} $\zeta^*_a(t)$, can be found as:
\begin{align}
\label{eq:pen_chance}
    \zeta^*_a(t) = \inf \left\{ \zeta \,\left|\, \Pr \left[ \bigcap_{k = 1}^K \left( \rho^k_a(t) + \phi^k_a \cdot \zeta \ge 0 \right) \right] \ge 1 - \epsilon \right. \right\},
\end{align}
\emph{i.e.}, the minimum demand that needs to be ``kicked out'' of $d_a(t)$ such that the remaining can be served without SLA.
With a penalty constant $P$, the system penalty is defined as
\begin{equation}
    \label{eq: penalty}
    P_a(t) = P \cdot \zeta^*_a(t).
\end{equation}

\textbf{EP's profit.}
The EP wants to maximize its profit $S(t)$:
\begin{equation}
    \label{fml:1}
    \max_{\mathbf{X}} \lim_{T \rightarrow \infty} \frac{1}{T} \! \sum_{t=1}^{T} \! {S(t)}, 
    \text{ where } S(t) \!\triangleq\! \sum\limits_a \! \left( U_a(t) \!-\! P_a(t) \right) \!-\! C(t).
\end{equation}

\section{URANUS Framework Overview}
\label{sec:overview}

\begin{figure}[htb]
\centering
{\includegraphics[width=0.45\textwidth, trim=150 190 150 190, clip]{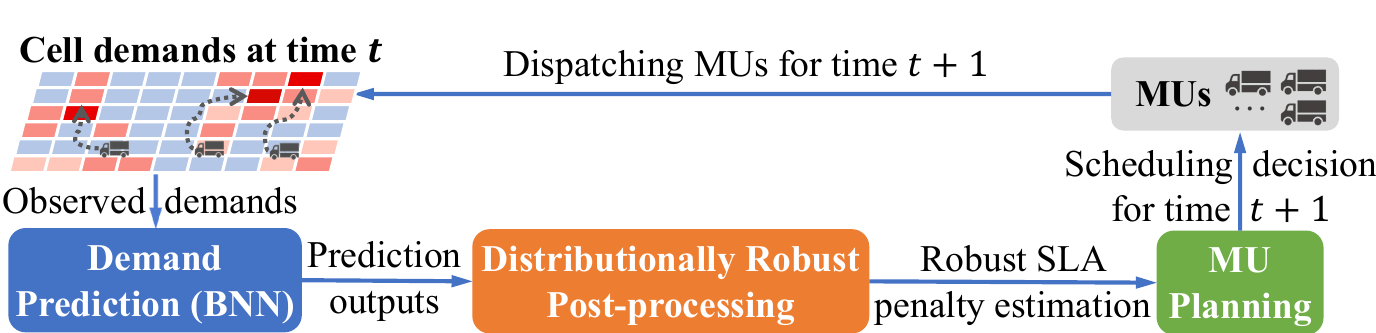}
}
\caption{The URANUS framework with three major components: BNN-based demand prediction, distributionally robust post-processing, and fixed-horizon MU planning}
\label{fig:uranus}
\end{figure}

\noindent
We design an online, learning-based MU scheduling framework for MOE-based on-demand edge computing as shown in Fig.~\ref{fig:uranus}, with a novel combination of uncertainty-aware learning, risk modeling, and deterministic planning.
Our framework is a generalized Model Predictive Control (MPC) algorithm that employs a predictive model of the system and environment to make control decisions.
At the end of each time $t$, a deep learning model makes a \textbf{fixed-horizon} prediction of future demands based on historical observations.
The EP then makes \textbf{robust estimation} of the expected utility and potential penalty for SLA violation.
Finally, a \textbf{fixed-horizon} planning algorithm decides the best action of each MU for time $t+1$.
We choose to design an MPC-type algorithm rather than directly using RL mainly because of the stability and robustness advantages of MPC in dealing with risk-related concerns such as SLA violation penalty.
Our overall design choice is motivated by:
\begin{itemize}[leftmargin=1em]
    \item \textbf{Predictive errors:} DNN models for time series prediction have lower accuracy over longer prediction spans. As we will show in Fig.~\ref{fig:blstm}, even a well-fitted model has large prediction errors (model uncertainty) when predicting multiple time steps into the future.
    \item \textbf{Fast dynamics:} Edge demands are highly fluctuating and uncertain, with high data uncertainty that cannot be addressed by even a most well trained model.
    \item \textbf{Excessive discounting:} RL employs a discount factor to exponentially reduce the weight of rewards of future steps. After a number of time steps, the weight will become negligible, leading to a procedure very similar to fixed-horizon planning as in MPC.
\end{itemize}

Combining these factors, instead of looking too deep into the uncertain future, an MPC-type algorithm tries to capture the most current trends within the most predictable near future, and makes optimized decisions based on a limited scope of well predicted information.
Formally, let $L_{\mathsf{pred}}$ be the length of the \emph{prediction horizon} on which a well trained model has empirically shown high enough accuracy. 
At the end of time $t$, the system predicts future demands within the horizon of $\omega_t = (t+1, t+2, \dots, t+L_{\mathsf{pred}})$, and then picks an action $x_i(t+1)$ for each MU $i$ based on the prediction over $\omega_t$.
\textbf{Challenges.}
Predictive models such as DNNs have been widely adopted in decision systems to dynamically predict future demands or rewards, including in DRL-based approaches.
However, these models have been reported to generally lack robustness against both \textbf{model} and \textbf{data uncertainties}~\cite{variational}, which has also been validated in our evaluation (Fig.~\ref{fig:blstm} in Sec.~\ref{sec:bnn}).
Decision making based on inaccurate information may lead to SLA violations and significant penalty.
Besides these, we also find that DNNs are generally ineffective in modeling complex forms of \emph{rewards} with problem-specific structures, and require enormous training data to achieve stable and reasonable learning in real-world. % DRL. 
Our evaluation confirms the above concerns on model-free RL approaches, motivating us to develop an MPC-based framework instead of using out-of-box DRL algorithms.

\section{Predicting Uncertain Edge Demands}
\label{sec:bnn}

\subsection{Bayesian Neural Network}
\newcommand{\dataset}{{{{{D}}}}}

\noindent
We model edge demand prediction of each cell $a \in \mathcal{A}$ as a time series prediction problem.
Consider an input horizon $\omega^{\mathsf{in}} = (t-L_{\mathsf{in}}+1, \dots, t)$ of length $L_{\mathsf{in}}$ denoting length of the input time series $\mathbf{d}^{\mathsf{in}} = \{ d_a(\tau) \,|\, \tau \in \omega^{\mathsf{in}} \}$.
The goal is to predict the demands $\mathbf{d}^{\mathsf{out}} = \{ d_a(\tau) \,|\, \tau \in \omega^{\mathsf{out}} \}$ over the prediction horizon $\omega^{\mathsf{out}} = \omega_t$.

Given a training dataset $\dataset{} = (\dataset{}^{\mathsf{in}}, \dataset{}^{\mathsf{out}})$, a canonical DNN tries to learn a parametric model, $h(\mathbf{d}^{\mathsf{in}} ; \theta)$, by minimizing the expectation of the empirical loss $L$:
\begin{equation}
\theta^* = \arg\min\nolimits_\theta \mathbb{E}_{(\mathbf{d}^{\mathsf{in}}, \mathbf{d}^{\mathsf{out}})\sim \dataset{}} \left[ L(h(\mathbf{d}^{\mathsf{in}} ; \theta), \mathbf{d}^{\mathsf{out}})\right ].
\end{equation}
The learned parameters $\theta^*$ is a \emph{point estimate} of the optimal DNN parameters for the training set.
Prediction of an unseen input $\mathbf{d}^{\mathsf{in}} \notin \dataset{}^{\mathsf{in}}$ is then $\mathbf{\widehat d}^{\mathsf{out}} = h(\mathbf{d}^{\mathsf{in}} ; \theta^*)$, which is also a \emph{point estimate} of the unknown true label $\mathbf{d}^{\mathsf{out}}$.
It has been shown in the literature that a point-estimate DNN can lead to issues such as overfitting and poor out-of-distribution performance, due to inability to consider model or data uncertainty~\cite{variational, bnn_survey}.

In a Bayesian neural network (BNN), a \emph{posterior distribution} $p(\theta \,|\, \dataset{})$ over model parameters is learned instead of a point estimate.
Based on the Bayes' theorem, given a prior distribution $p(\theta)$ representing initial belief, the posterior can be derived as: 
\begin{equation}
\label{eq:bnn}
p(\theta \,|\, \dataset{}) \!=\! \frac{
    p(\dataset{}^{\mathsf{out}} \,|\, \dataset{}^{\mathsf{in}}, \theta) p(\theta)
}{ 
    \int_{\Tilde{\theta}} p(\dataset{}^{\mathsf{out}}  \,|\, \dataset{}^{\mathsf{in}}, \Tilde{\theta}) p(\Tilde{\theta}) d\Tilde{\theta}
} \propto p(\dataset{}^{\mathsf{out}} \,|\, \dataset{}^{\mathsf{in}}, \theta) p(\theta).
\end{equation}
A prediction on $\mathbf{d}^{\mathsf{in}}\notin \dataset{}^{\mathsf{in}}$ is made by combining the posterior distribution with the likelihood function into the \emph{marginal distribution}:
\begin{equation}
\label{eq:marginal}
p(\mathbf{d}^{\mathsf{out}} \,|\, \mathbf{d}^{\mathsf{in}}, \dataset{}) = \int_\theta p(\mathbf{d}^{\mathsf{out}} \,|\, \mathbf{d}^{\mathsf{in}}, \theta) p(\theta \,|\, \dataset{}) d\theta.
\end{equation}

Prediction in the form of a marginal distribution integrates both data and model uncertainties~\cite{uncertainty_review}.
Specifically, the posterior $p(\theta \,|\, \dataset{})$ embeds model uncertainty on the training dataset, and the likelihood $p(\mathbf{d}^{\mathsf{out}} \,|\, \mathbf{d}^{\mathsf{in}}, \theta)$ embeds data uncertainty each model faces when predicting for a new input $\mathbf{d}^{\mathsf{in}}$.

Meanwhile, BNN in the form of~\eqref{eq:bnn} is intractable to train and sample from for complex DNN models, due to the high dimensionality and non-convexity of the posterior as well as the difficulty in computing the integral $\int_{\Tilde{\theta}} p(\dataset{}^{\mathsf{out}}  \,|\, \dataset{}^{\mathsf{in}}, \Tilde{\theta}) p(\Tilde{\theta}) d\Tilde{\theta}$.

\subsection{Variational BNN for Demand Prediction}

\noindent
Since~\eqref{eq:bnn} is intractable for neural networks, we use a variational approximation~\cite{variational} to the posterior distribution based on Kullback-Leibler (KL) divergence and Monte Carlo approximation.
Consider the posterior $p(\theta | D)$ being approximated by a parametric distribution $p(\theta | \mathbf{w})$ parameterized by $\mathbf{w}$.
The training objective is:
\begin{equation}
    \mathbf{w}^* = \arg \min_{\mathbf{w}} KL\left[p(\theta|\mathbf{w}) || p(\theta|D)\right].
\end{equation}
Variational learning aims to find the parameter $\mathbf{w}$ that minimizes the KL divergence between the approximate posterior $p(\theta | \mathbf{w})$ and the true posterior $p(\theta|D)$.
To train the variational model, we adopt a training algorithm similar to stochastic gradient descent (SGD) based on Monte Carlo sampling~\cite{variational}.

\textbf{Training.}
Consider $\mathcal{T}_{\mathsf{train}} \!=\! \{ 1, 2, \dots, L_{\mathsf{train}} \}$ time steps of historical demands $\{ d_a(t) \,|\, {t \in \mathcal{T}_{\mathsf{train}}} \}$ for cell $a$.
Each training sequence consists of $(L_{\mathsf{in}} + L_{\mathsf{pred}})$ consecutive time steps of demand values, with the first $L_{\mathsf{in}}$ as input features and the last $L_{\mathsf{pred}}$ as input labels.
Training dataset $D$ thus contains $L_{\mathsf{train}}-(L_{\mathsf{in}} + L_{\mathsf{pred}})+1$ sequences in total.
These sequences are regarded as IID samples to train the BNN model.
We refer the reader to~\cite{variational} for the BNN training algorithm.

\textbf{Prediction.}
For real-time prediction at time $t \in \mathcal{T}$, input is the last $L_{\mathsf{in}}$ time steps of historical demands $\mathbf{d}^{\mathsf{in}}_a(t) \!=\! (d_a(t-L_{\mathsf{in}}+1), \dots, d_a(t))$.
The model is sampled for multiple times.
Each sample takes $\mathbf{d}^{\mathsf{in}}_a(t)$ as input and outputs a point estimate of label $\mathbf{d}^{\mathsf{out}}_a(t)$.
Two $L_{\mathsf{pred}}$-dimension vectors $\bm{\mu}_a(t) = ( \mu_{a, {{{\tau}}}}(t) )_{{{{\tau}}} \in \omega_t}$ and $\bm{\sigma}_a(t) = ( \sigma_{a, {{{\tau}}}}(t) )_{{{{\tau}}} \in \omega_t}$ are then computed as mean and standard deviation of the sample outputs, which respectively estimate the mean and standard deviation of the predicted uncertain demand distribution for each time step ${{{\tau}}}$ in $\omega_t$.
For simplicity of notations, let us use $\bm{\mu}(t) = \{ 
\bm{\mu}_a(t) \,|\, a \}$ and $\bm{\sigma}(t) = \{ 
\bm{\sigma}_a(t) \,|\, a \}$ to denote all the predictions made at time $t$.
%
% %
Assuming the marginal distribution is approximated by a Gaussian distribution,
the $95\%$-confidence interval (CI) can be computed as $\mathsf{CI}^{0.95}(t) \approx \bm{\mu}(t) \pm 2 \cdot \bm{\sigma}(t)$.

\subsection{Distributional Uncertainty in BNN}
\label{sec:dist_uncert}

\begin{figure}[htb]
\centering
{\includegraphics[width=0.47\textwidth, trim=170 140 160 70, clip]{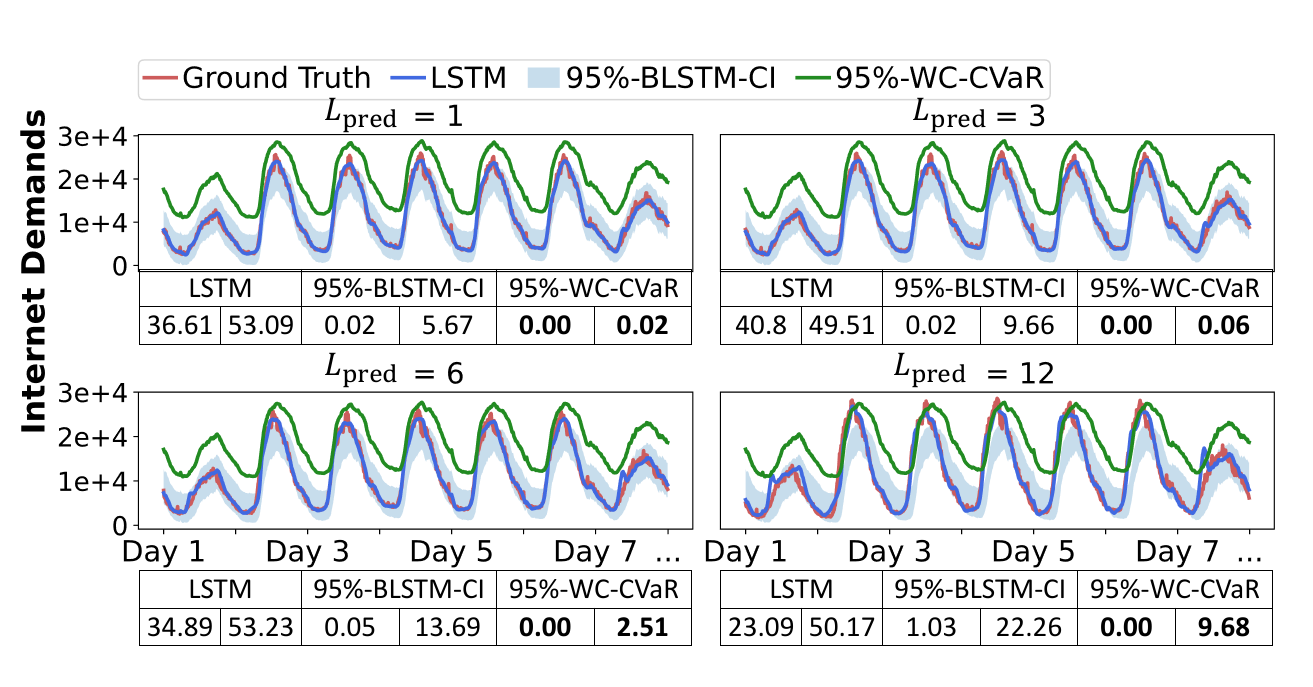}
}
\caption{Prediction outputs of LSTM, BLSTM 95\%-CI, 95\% WC-CVaR are shown in sub-figures under different $L_{\sf pred}$. Tables below sub-figures show the \emph{best} and \emph{worst} percentages of prediction violations among all cells.}
\label{fig:blstm}
\end{figure}

\noindent
Existing works applying BNN have mostly assumed it can perfectly capture uncertainties.
However, recent research~\cite{izmailov2021bayesian, wenzel2020good} shows that BNN is subject to inaccurate uncertainty quantification due to various approximations used.
In demand prediction, we make the same observation on real-world data, which can be due to a mixture of limited training data, long time series horizon to predict, variational approximation, as well as difficulty in obtaining sufficient Monte Carlo samples to estimate the CI at a high confidence level.
Specifically, we trained a Bayesian Long Short-Term Memory (BLSTM) model on $15$ days of telecommunication data in Milan~\cite{mi_data}, preprocessed into $20,050$ sequences with $L_{\mathsf{in}} \!=\! 144$ and $L_{\mathsf{pred}} \! \in \! \{ 1, 3, 6, 12 \}$.
Details of the trained models are in Sec.~\ref{sec:eval}.
Fig.~\ref{fig:blstm} shows demand prediction for the next $7$ days, where we compare the upper $95\%$-CI based on BLSTM output samples (with the Gaussian assumption), to the ground truth and point-estimate output of a (non-Bayesian) LSTM with the same dimensions.
The percentage of ``violations'', defined as a predicted value being \emph{exceeded} by the ground truth (cases where resources may be insufficiently provisioned), is shown below each figure.
Three critical observations motivate our design:
1) While LSTM generally fits better with the ground truth, it actually leads to much more violations than $95\%$-BLSTM-CI, motivating the use of BLSTM.
2) Even for BLSTM, the fraction of violations is still notably larger than the calculated CIs, \emph{i.e.}, there are much more than $5\%$ cases where the ground truth exceeds the $95\%$-CI.
This indicates \emph{miscalibrated uncertainty quantification} due to \textbf{distributional uncertainty}, \emph{i.e.}, uncertainty in the posterior/marginal distribution, and it motivates our design with the WC-CVaR risk measure (explained in Sec.~\ref{sec:risk}).
3) Violations increase with larger $L_{\mathsf{pred}}$, motivating the finite-horizon planning in Sec.~\ref{sec:dp}.%

\section{Solving Distributional Uncertainty}
\label{sec:risk}

\noindent
To tackle distributional uncertainty of BNN outputs, the outputs need to be post-processed for robust decision making at time $t$.
This section addresses robustness via stochastic optimization.
In this and the next sections, we omit current time $t$ in symbols whenever appropriate, \emph{e.g.},
we denote BNN outputs as $\bm{\mu}_a$ and $\bm{\sigma}_a$ for each $a$.

\subsection{Robust Stochastic Optimization}
\noindent
Recall that EP's profit consists of two random variables, utility $U_a$ and penalty $P_a$, whose expectations need to be estimated based on BNN outputs.
Estimating $\mathbb{E}[U_a]$ is simple, as one can apply \emph{sample average approximation} based on BNN output samples, inputting samples into the utility function and calculating the mean.

The difficulty lies in estimating $\mathbb{E}[ P_a ]$, defined upon the expected excess demand $\zeta^*_a$, which is the optimal value of a chance-constrained stochastic program in Eq.~\eqref{eq:pen_chance}.
This is difficult because: 1) we do not have a closed form of either the true or predicted distribution of $d_a$, but only predicted samples and statistics ($\boldsymbol{\mu}_a, \boldsymbol{\sigma}_a$); 2) the predicted distribution may be shifted from the true distribution as we have shown; 3) the chance constraint in Eq.~\eqref{eq:pen_chance} is neither differentiable nor convex.
We address the combined challenge with a \emph{distributionally robust approximation} of the chance constraint.

The \emph{value-at-risk (VaR)} and \emph{conditional value-at-risk (CVaR)} are two risk measures widely used in economics and finance~\cite{optimization_cvar}. % as also indicted in~\cite{GC21, MOBIHOC20, CNS18}, 
Given random variable $R$ denoting the investment loss with probability density function (PDF) $\mathbb{P}$ and cumulative density function (CDF) $\mathbb{F}$, they are defined as:
\begin{align}
\label{eq:cvar}
    & \mathbb{P}\text{--VaR}_{\alpha}(R) \triangleq \inf \{ \pi \,|\, \mathbb{F}[ \pi ] \ge \alpha \}; \\
    & \mathbb{P}\text{--CVaR}_{\alpha}(R) \triangleq \mathbb{E}_\mathbb{P}[ R \,|\, R \ge \mathbb{P}\text{--VaR}_{\alpha}(R) ].
\end{align}
Here $\mathbb{P}\text{--VaR}_{\alpha}(R)$ indicates the threshold value $\pi$ of the investment loss, such that with at least $\alpha$ probability, the actual loss does not exceed $\pi$.
$\mathbb{P}\text{--CVaR}_{\alpha}(R)$ then reflects the expected investment loss when the loss actually exceeds the given threshold $\pi$.
Note that the definition of CVaR depends on VaR.
Let $\alpha = 1-\epsilon$.
The following property holds for these two measures.
\begin{property}
    \label{prop:cvar}
    $\mathbb{P}\text{--CVaR}_{\alpha}(R) \ge \mathbb{P}\text{--VaR}_{\alpha}(R)$.
\end{property}

To facilitate further formulation, we define the \emph{excess demand} at cell $a$ as follows (based on definition of $\mathbf{f}_a(\cdot)$ in Sec.~\ref{sec:demand_model}): 
\begin{equation}
    \label{eq:excess}
    \zeta_a = \max_k\left\{ \frac{-\rho^k_a}{\phi^k_a} \right\} = %\min \left\{
        \max_k\left\{ \frac{f_a^k(d_a) - r_a^k - g^k \cdot z_a}{\phi_a^k} \right\}. 
    %, d_a \right\}.
\end{equation}
This transforms the $k$ random variables $\rho^k_a$ into one single random variable $-\zeta_a(t)$ based on the demand $d_a$.
One can verify that given constant $\zeta$, $\bigcap_{k = 1}^K ( \rho^k_a + \phi^k_a \cdot \zeta \ge 0 )$ whenever $\zeta_a - \zeta \le \! 0$, and vice versa.
Hence Eq.~\eqref{eq:pen_chance} can be re-written as:
\begin{align}
\label{eq:pen_chance1}
    \zeta^*_a = \inf \left\{ \zeta \,\left|\, \Pr \left[ \zeta_a - \zeta \le 0 \right] \ge 1 - \epsilon \right. \right\} = \mathbb{Q}\text{-VaR}_{\alpha}(\zeta_a),
\end{align}
where $\mathbb{Q}$ is the PDF of $\zeta_a$.
Because VaR is non-convex and non-differentiable, we utilize Property~\ref{prop:cvar} and make a conservative approximation of the VaR by using the CVaR which is convex~\cite{cvar}:
\begin{align}
\label{eq:pen_chance2}
    \zeta^*_a \approx \mathbb{Q}\text{-CVaR}_{\alpha}(\zeta_a).
\end{align}

\subsection{Distributionally Robust Approximation}

\noindent
Directly using Eq.~\eqref{eq:pen_chance2} to estimate $\zeta^*_a$ and hence $P_a$ still has two issues:
1) the definition of CVaR depends on VaR which is non-convex;
2) it assumes knowledge of the distribution $\mathbb{Q}$ of $\zeta_a$ which depends on the true distribution of demand $d_a$.
Yet, we only have an approximation to the distribution of $d_a$ based on the BNN output, with potential distributional uncertainty.
Here, we are motivated to take a robust approach, by considering \emph{all possible distributions with the same first- and second-order moments} as the marginal samples.

Consider a time step within the current prediction horizon: ${{{\tau}}} \in \omega_t$.
Define ambiguity set $\mathcal{Q}$ as the set of all distributions with mean $\mu_{a, {{{{\tau}}}}}$ and variance $\sigma_{a, {{{{\tau}}}}}^2$.
We have the following \textbf{worst-case CVaR} (WC-CVaR) estimation of $\zeta^*_{a, {{{\tau}}}}$ for ${{{\tau}}}$:
\begin{align}
\label{eq:cvar-cons-suff}
    \zeta^*_{a, {{{\tau}}}} = \sup\nolimits_{\mathbb{Q} \in \mathcal{Q}} \mathbb{Q}\text{--CVaR}_{1-\epsilon}( \zeta_{a, {{{{\tau}}}}}),
    \;
    \forall a, {{{{\tau}}}}.
\end{align}
$\zeta_{a, {{{{\tau}}}}}$ is an approximation to $\zeta_{a}$ at time $\tau$, based on any possible predictive distribution $\mathbb{Q}$ within the uncertainty set $\mathcal{Q}$.

\textbf{Remark:}
Considering that the BNN marginal distribution can be ill-shaped (non-Gaussian) due to variational approximations applied, the intuition behind WC-CVaR is to robustly estimate the excess demand directly with the first- and second-order moments of the marginal samples.
This provides notably improved robustness than assuming $\mathbb{Q}$ is Gaussian.
To illustrate why this would work, in Fig.~\ref{fig:blstm} we also show the WC-CVaR of demand prediction as compared to the pure LSTM or $95\%$-BLSTM-CI predictions.
To make a direct comparison, we slightly modified the definition to directly define WC-CVaR on the demand instead of excess demand, different from Eq.~\eqref{eq:cvar-cons-suff}.
We observe that, the WC-CVaR with $\alpha \! = \! 95\%$ leads to notably fewer violations than the $95\%$-CI even when $L_{\sf pred} \! = \! 12$, and the violations are bounded by the desired percentage ($5\%$) in most cases.
Assuming CDF of the true distribution is upper bounded by CDF of any $\mathbb{Q} \in \mathcal{Q}$, Eq.~\eqref{eq:cvar-cons-suff} is an upper bound of $\zeta^*_{a, \tau}$, leading to a conservative estimation of $P_a$.

\subsection{Tractable Reformulation with SDP}

\noindent
We finally address tractability of~\eqref{eq:cvar-cons-suff}.
The so-defined $\zeta_{a, {{{{\tau}}}}}^*$ is still complicated by the infinite number of distributions in $\mathcal{Q}$, as well as non-convexity of the VaR metric which appears in the definition of CVaR.
In Theorem~\ref{th:sdp}, we show that $\zeta_{a, {{{{\tau}}}}}^*$ equals the optimal value of a \emph{semi-definite program (SDP)} inspired by the technique in~\cite{cvar}.
\begin{theorem}
\label{th:sdp}
Consider a cell $a$ with predictive sample mean $\mu_{a, {{{{\tau}}}}}$ and standard deviation $\sigma_{a, {{{{\tau}}}}}$ of the demand $d_a({{{\tau}}})$. 
Let $z_{a, {{{\tau}}}}$ be the number of MUs scheduled to be operational in cell $a$ at time ${{{\tau}}}$.
The estimated expected excess demand $\zeta_{a, {{{{\tau}}}}}^*$ can be computed by 
\begin{subequations}
\label{fml:sdp}
\begin{align}
&\zeta_{a, {{{{\tau}}}}}^* = \min {{{\xi}}}
\quad
\text{ \sf s.t. } {{{\xi}}} \in \mathcal{Y},
\tag{\ref{fml:sdp}} 
\\
\label{sdp}
&    \mathcal{Y} = 
    \left\{  
    {{{\xi}}} \in \mathbb{R} \,\left|\, 
    \begin{array}{l}
        \exists {{{\nu}}} \in \mathbb{R}, \mathbf{M} \in \mathbb{S}^2, {{{\xi}}}^k \in \mathbb{R} \;\; \forall k, \text{ such that } \\
        {{{\nu}}} + \frac{1}{\epsilon} \langle \mathbf{\Omega}, \mathbf{M} \rangle \le 0; \;  \mathbf{M} \succeq 0; \\
        \forall k:
        \text{for simplicity, let}  \\
        % \Gamma^k_a(t) =  \frac{\varphi^k_a(t) - r^k_a }{g^k} - z_a(t) - {{{\nu}}} - \phi^k_a \zeta^k, \\
        \Gamma^k =  \frac{\varphi^k_a - r^k_a - g^k z_{a, {{{\tau}}}} }{\phi^k_a} - {{{\nu}}} - {{{\xi}}}^k, \text{then} \\
        \mathbf{M} - \left[ \begin{array}{cc}
            0   
            & 
            {1}/{2} 
            \\
            {1}/{2} 
            & 
            \Gamma^k
        \end{array}\right] \succeq 0, \text{ and } 
        {{{\xi}}} \ge {{{\xi}}}^k;
    \end{array}
    \right.
    \right\},
\end{align}
\end{subequations}
where $\mathbb{R}$ is the real number set, $\mathbb{S}^2$ is the set of $2\times 2$ symmetric matrices, and $\mathbf{\Omega}$ is a second-order matrix defined as 
$\mathbf{\Omega} = 
\left[ 
\begin{array}{cc}  
\mu_{a, {{{{\tau}}}}}^2 + \sigma_{a, {{{{\tau}}}}}^2 & \mu_{a, {{{{\tau}}}}} \\ 
\mu_{a, {{{{\tau}}}}} & 1  
\end{array} 
\right].$
\end{theorem}
The set $\mathcal{Y}$ above defines a polynomial set of semi-definite constraints on auxiliary variables $\mathbf{M}$, $\nu$, ${{{\xi}}}^k$ and ${{{\xi}}}$ (defined only within the scope of this program).
The minimum value $\xi$ gives the WC-CVaR value $\zeta^*_{a, {{{{\tau}}}}} \triangleq \sup\nolimits_{\mathbb{Q} \in \mathcal{Q}} \mathbb{Q}\text{--CVaR}_{1-\epsilon}( \zeta_{a, {{{{\tau}}}}} )$ given scheduling decision $z_{a,{{{\tau}}}}$ and 
predicted $\mu_{a, \tau}$, $\sigma_{a, \tau}$.
The proof is an extension from~\cite{cvar} and is omitted due to page limit.
Theorem~\ref{th:sdp} allows us to compute the expected penalty $P_{a, {{{{\tau}}}}} = P \cdot \max\{ \zeta^*_{a, {{{\tau}}}}, 0 \}$ given decision $z_{a, {{{{\tau}}}}}$. 
Consider that we want to make \emph{tentative} scheduling decisions for $\tau \in \omega_t$.
Let $\mathbf{X}_{\{\tau\}} = \{ \mathbf{X}_{\tau} \,|\, \tau \in \omega_t \}$ be \emph{tentative decision variables}, and $U_{a, \tau}, P_{a, \tau}, C_\tau$ be the \emph{tentative utility, penalty and cost} respectively for $\tau \in \omega_t$.
Our one-shot problem at time $t$ is to maximize the expected EP profit within $\omega_t$:
\begin{equation}
\label{fml:3}
    \text{(P2)}
    \quad
    \max_{
    \mathbf{X}
    }\;
    \sum_{ \tau \in \omega_t } \left(
    \sum_{a\in \mathcal{A}} \left(U_{a, {{{\tau}}}} - P_{a, {{{{\tau}}}}} \right) - C_\tau) \right).
\end{equation}

\section{{Planning for a Finite Horizon}}
\label{sec:dp}

\noindent
Given prediction horizon $\omega_t$, a one-shot planning algorithm seeks to pick an action for each MU to take in time step $t+1$.
However, the algorithm needs to consider impact of future actions.
Below, we develop a single-agent optimal planning algorithm, and then extend it greedily for multiple agents.

\subsection{Single-Agent Planning}

\begin{figure}[htb]
\centering
{\includegraphics[width=0.43\textwidth, trim=140 140 120 140, clip]{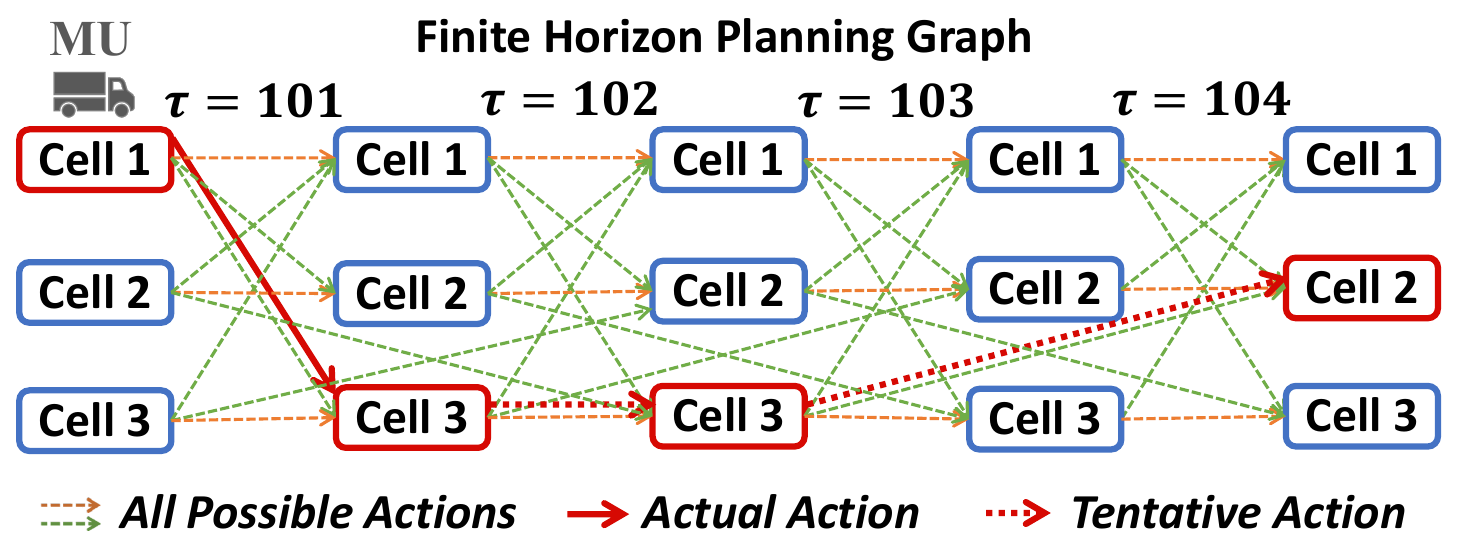}
}
\caption{Planning for three cells at $t = 100$ when ${L_{\mathsf{pred}}}=4$}
\label{fig:dp}
\end{figure}

\noindent
Consider there is only one MU to schedule.
We can model trajectory of the MU in horizon $\omega_t$ as a \emph{path}, as shown in Fig.~\ref{fig:dp}.
In each time $\tau \in \omega_t$, the MU chooses to move to a different cell, or to stay in the current cell; if staying, it chooses to become operational and serve user demands, or stay idle and save cost.
After each time step, the system receives reward (profit) from all cells, which depends on the actual demand, the fixed resources, and operation of the MU.

Assume the MU is in cell $a \in \mathcal{A}$ at the end of time $\tau-1$.
Let $U^0_{a, {{{\tau}}}}$ and $U^1_{a, {{{\tau}}}}$ be the expected utility from cell $a$ at time $\tau$, when $0$ and $1$ MU is scheduled to be operational in $a$ respectively.
Similarly define penalty $P^0_{a, {{{\tau}}}}$ and $P^1_{a, {{{\tau}}}}$.
The \emph{reward gain} for the MU to be \emph{operational} in location $a$ at time $\tau$ is $\Delta S_{a, \tau}^{\mathsf{opt}, 1} = U^1_{a, {{{\tau}}}} + P^1_{a, {{{\tau}}}} - U^0_{a, {{{\tau}}}} - P^0_{a, {{{\tau}}}} - c_{\mathsf{opt}}$, where the superscript $(\mathsf{opt}, 1)$ means having $1$ versus $0$ MU scheduled to be operational in $a$.
The \emph{reward gain} for the MU to be \emph{idle} is $\Delta S_{a, \tau}^{\mathsf{idl}} = -c_{\mathsf{idl}}$.
The \emph{reward gain} for the MU to transit to a remote cell $b$ is $\Delta S_{ab, {{{\tau}}}}^{\mathsf{tra}} = -c_{\mathsf{tra}} \cdot T_{ab}$.
Note that the \emph{in transit} cost is counted for the next $T_{ab}$ steps taken to complete transit to $b$, during which the MU is not schedulable.
Based on these, we can assign a value to each possible state-action tuple $(a_{\tau-1}, x_{\tau}, \delta_{\tau}) \in \{ (a, \mathsf{opt}, \perp), (a, \mathsf{tra}, b), (a, \mathsf{idl}, \perp) \,|\, \forall a, b \in \mathcal{A} \}$ where $a_{\tau-1}$ is the state (location) at the end of $\tau-1$ and $(x_{\tau}, \delta_{\tau})$ is the action (mode and new location).

To build a graph $G = (V, E)$ as in Fig.~\ref{fig:dp}, define node set $V = \{ (a, \tau) \,|\, a \in \mathcal{A}, \tau \in \{ t \} \cup \omega_t \}$ where $t$ is current time, and link set $E = E_{\mathsf{stay}} \cup E_{\mathsf{tra}}$ where $E_{\mathsf{stay}} = \{ ((a, \tau), (a, \tau+1)) \,|\, a, \tau \}$ and $E_{\mathsf{tra}} = \{ ((a, \tau), (b, \tau + T_{ab}))\,|\, a, b, \tau \}$.
For each link $e = ((a, \tau), (a, \tau+1)) \in E_{\mathsf{stay}}$, define its \emph{length} as $l_{e} = \min\{ -\Delta S_{a, \tau}^{\mathsf{opt}}, -\Delta S_{a, \tau}^{\mathsf{idl}} \}$.
For each link $e = ((a, \tau), (b, \tau+T_{ab})) \in E_{\mathsf{tra}}$, define its \emph{length} as $l_{e} = -\Delta S_{ab, {{{\tau}}}}^{\mathsf{tra}}$.
The total length $l_p \!=\! \sum_{e \in p} l_e$ is precisely the \emph{negative} total reward gain of this entire path.
Let $a(t)$ be location of the MU at the end of time $t$.
Maximizing the reward within horizon $\omega_t$ is equivalent to finding a \emph{minimum-length path} in $G$, starting from node $(a(t), t)$ to any node $(b, t + L_{\mathsf{pred}})$.
Since $G$ is a \emph{directed acyclic graph (DAG)}, a minimum-length path can be found in $O(|V| + |E|) \!=\! O(|\mathcal{A}|^2 L_{\mathsf{pred}})$ time~\cite{algo_book}. 
The running time per-time step single-agent planning is thus $t_B+ L_{\sf pred}\cdot|\mathcal{A}|\cdot t_S+ t_P$, where $t_B$ is BNN prediction time (constant), $t_S$ is single SDP solver time (constant) and $t_P=|\mathcal{A}|^2 L_{\mathsf{pred}}$ is the planning time.

\subsection{Multi-Agent Planning}
\noindent
Planning for multiple MUs is more involved, as scheduling of one MU affects both its own \emph{reward gain} and the \emph{reward gain} for scheduling other MUs.
This impact is non-linear as both utility and penalty can be non-linear, and the utility in Eq.~\eqref{eq: utility} is non-convex in number of operational MUs.
Instead, we rely on a greedy algorithm where MUs are scheduled one-by-one over the horizon $\omega_t$, each based on the scheduling decisions of all previous MUs.
For any MU scheduled to be \emph{operational} in cell $a$ at time $\tau$, it contributes $\mathbf{g}$ resources to the cell, which is regarded as \emph{fixed resources} when scheduling subsequent MUs.
Accordingly, when scheduling the $i$-th MU, we can compute the \emph{reward gain}
$\Delta S_{a, \tau}^{\mathsf{opt}, i} = U^{z_{a, \tau}+1}_{a, {{{\tau}}}} + P^{z_{a, \tau}+1}_{a, {{{\tau}}}} - U^{z_{a, \tau}}_{a, {{{\tau}}}} - P^{z_{a, \tau}}_{a, {{{\tau}}}} - c_{\mathsf{opt}}$ based on current $z_{a, \tau}$ as decided by MUs $1, 2, \dots, i-1$.
Given an arbitrary order of scheduling among the MUs, the total complexity per time step for multi-agent planning has three terms: $t_B + m \cdot (L_{\sf pred} \cdot |\mathcal{A}| \cdot t_S + t_P) $, where $m$ is number of MUs, $L_{\sf pred}$ is the prediction horizon and $|\mathcal{A}|$ is the number of cells.
We note that instead of scheduling in an arbitrary order, the MUs can also be scheduled following a greedy order such as always scheduling the MU resulting in the largest reward gain first.
However, our evaluation shows that the scheduling order has negligible impact, while employing the greedy order requires an order of magnitude higher time complexity.
\begin{algorithm}[t]
\caption{URANUS Workflow}
\label{a:uranus}
    \KwIn{Cells $\mathcal{A}$ with resources $\{\mathbf{r}_a\}$, number of MUs $m$, per-MU resources $\mathbf{g}$, pre-trained BNN model $h_B$ with parameters $\mathbf{w}^*$.}
    \KwOut{ Scheduling decisions $\mathbf{X}(t)$ for $t = 1, 2, \dots$. }
    \For{time step $t = 1, 2, \cdots$}{
        Update current MU locations based on initial location, or decisions in the previous time step(s)\;
        \tcp{Demand Prediction Phase}

        \For{$a \in \mathcal{A}$} {
            Get actual demands $\mathbf{d}^{\mathsf{in}}_a(t)$ of $L_{\mathsf{in}}$ past time steps\;
            $\bm{\mu}_a(t), \bm{\sigma}_a(t) \gets h_B(\mathbf{d}^{\mathsf{in}}_a(t); \mathbf{w}^*)$\;
        }
        
        \tcp{MU Scheduling Phase}
        \For{MU $i = 1, 2, \dots, m$}{
            \For{$a \in \mathcal{A}$ and $\tau \in \omega_t$}{
                Compute $\Delta S^{\mathsf{opt}, i}_{a, \tau}$ and $\Delta S^{\mathsf{idl}}_{a, \tau}$ for cell $a$\;
                Compute $\Delta S^{\mathsf{tra}}_{ab, \tau}$ for $\forall b \in \mathcal{A} \setminus \{ a \}$\;
            }
            Construct scheduling graph $G = (V, E)$\;
            Find minimum-length path $p$ in $G$ from $(a_i(t), t)$ to $(b, t+L_{\mathsf{pred}})$ among any $b \in \mathcal{A}$\;
            Set MU $i$'s action as indicated by the \textbf{first link} on $p$.\ \label{line:execute}
        }
    }
\end{algorithm}

\subsection{{URANUS Workflow}}

\noindent
Define $a_i(t)$ as location of MU $i$ at the end of time $t$.
The overall workflow is in Algorithm~\ref{a:uranus}.
Decision making for each time step consists of two phases: 1) using the BNN to predict the demands, 2) scheduling MUs based on the prediction.

One key note in the framework is that, while the prediction and planning horizon consists of $L_{\mathsf{pred}}$ time steps in the future, only actions planned for the \textbf{immediate next step}, $t+1$, are \emph{actually} executed by the MUs.
This corresponds to the \emph{first link} along the scheduling path $p$, as shown in Line~\ref{line:execute}.
This is because as time passes by, the actual demands may not perfectly match with the predicted demands, and dynamic adjustment to the scheduling needs to be done per-time step to ensure decision making with the most up-to-date information.
Meanwhile the other ``tentative actions'' are also essential, as they indicate the \emph{cumulative reward gain} that an agent can receive by taking a specific action at time $t+1$.
Without this ``looking into the future'', the scheduling algorithm will be based on only predicted demands in the next step, and loses the ability to \emph{plan} for surging hotspots in the future.

\subsection{Discussions}

\noindent
We have outlined the elements of the URANUS framework.
Here, we discuss practical implementation considerations.

\textbf{Reducing time complexity.}
While asymptotically $t_P \!=\! O(|\mathcal{A}|^2 L_{\sf pred})$ is the largest term in the time complexity, in evaluation, the dominant complexity comes from SDP solving time $m \cdot L_{\sf pred} \cdot |\mathcal{A}| \cdot t_S$ which is orders of magnitude larger than both $t_B$ and $m\cdot t_P$ due to $t_S$ being a large constant running time even with a small $k\!=\!3$.
To reduce the time complexity for solving many SDPs per planning (one per predicted time step per cell), the SDP may be replaced by an approximation function for approximating the robust outputs of WC-CVaR.
While finding the best WC-CVaR approximation is beyond our scope, in evaluation, we implemented a multi-layer perceptron (MLP) to efficiently predict the WC-CVaR values, pre-trained on SDP input-outputs. 
Our evaluation show that the MLP approximator can notably improve decision making efficiency, while having some trade-off on its performance.

\textbf{Applying a discount factor.}
It is generally assumed that the MPC model provides an accurate prediction of the system's behavior.
However, as in Fig.~\ref{fig:blstm}, we observe drastically degrading prediction accuracy when predicting further into the future (with a larger $L_{\sf pred})$.
Therefore, it is reasonable to apply a discount factor $\eta$ similarly as in infinite-horizon RL to counter this effect.
This also makes a somewhat fair comparison with RL.
We also empirically find that $\eta$ has almost negligible impact with a relatively small $L_{\sf pred}$, and the degrading prediction accuracy can be dealt with by picking the proper $L_{\sf pred}$ as well.
We apply $\eta=0.8$ in our evaluation.

\section{Performance Evaluation}
\label{sec:eval}

\subsection{Experiment Setup}

\subsubsection{Dataset and models}
\noindent
We used the read-world Internet activity traffic between November and December 2013 in Milan~\cite{mi_data}. 
\begin{figure}[b]
\centering
{\includegraphics[width=0.45\textwidth, trim=60 170 100 210, clip]{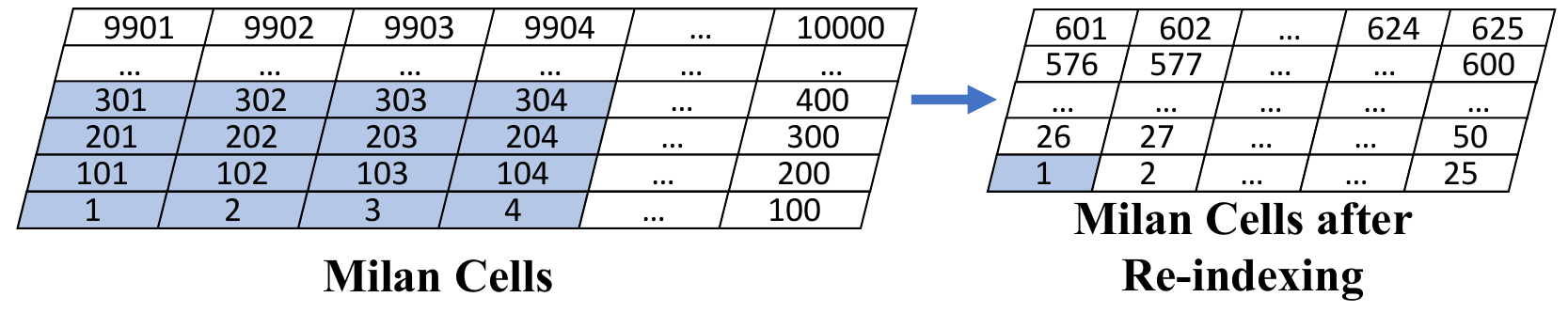}
}
\caption{Cell re-indexing for Milan cells~\cite{mi_data}}
\label{fig:grid}
\end{figure}
The area of Milan was divided into $10000$ cells of $235\text{m} \times 235\text{m}$ area as in Fig.~\ref{fig:grid}.
We merged and re-indexed these cells into larger $940\text{m} \times 940\text{m}$ cells to simulate the typical service area of a 5G base station.
%
%=
$10$ cells with maximum average demands were selected as $\mathcal{A}$ to represent real-world hotspots. 
Each time step consisted of $10$ minutes.

We divided the 2-month data $\mathcal{D}$ into 3 parts: $D_1$ (the first $15$ days of data), $D_2$ (the second $15$ days), and $D_3$ (the $31$ days left).
The prediction input $L_{\mathsf{in}} \!=\! 144$ (1 day) and output $L_{\mathsf{pred}} \!=\! 12$ (2 hours) by default.
For fair comparison, $D_1$ was used to train the BNN with time series sequences from top-10 cells, $D_2$ was used to train  the MLP for learning SDP solver as well as the RL for learning policy, and $D_3$ was used as test dataset for evaluation.
Since all picked hotspots were in urban areas, we estimated transit times based on distances between cell centroids, with MU's speed of $15$ km/h (considering possible delays).
It took maximally $2$ time steps to transit between cells.

For the BNN model, we trained a $1$-layer BLSTM with $50$ hidden states followed by a fully connected layer, with $42,212$ parameters.
A non-Bayesian LSTM of the same dimensions was trained for comparison, with $11,212$ parameters.
The MLP as SDP learner was a $5$-layer MLP with $44,289$ parameters.
The BLSTM/LSTM models were trained for $400$ epochs with batch size $b\!=\!128$, and the MLP for $500$ epochs with $b\!=\!64$.
The learning rate was $5\mathrm{e}{-4}$ for BLSTM and $1\mathrm{e}{-3}$ for LSTM/MLP.
The default penalty constant $P\! =\! 5\mathrm{e}{+5}$.

Three types of resources (CPU, GPU, memory) were configured for our setting.
$\phi_a \!=\! \left[ 0, 0, 10/30 \right]$ and $\varphi_a \!=\! \left[ 8, 1, 8 \right]$ are set as the coefficient and constant parameters for the CPU, GPU and memory resources, changing the memory coefficient between $10$ and $30$ for day and night hotspots respectively.
CPU and GPU minimum resources were set as constant to the demand ($\phi_a^k = 0$).
We set $r_a^k \!=\! \left[ 16, 2, 24\right]$ as the static resources at each cell while $g^k \!=\! \left[ 16, 2, 32\right]$ denoted the resources every MU carried.
The setting satisfied that the fixed resources in each cell could cover up the constant minimum resources to boot up all services, but might not be able to handle memory consumption with increasing user demand.

\subsubsection{Comparison algorithms}
\noindent
Our full URANUS algorithm combining BLSTM, SDP and planning was called \textbf{SP}.
When MLP was used as a learner of the SDP solver, we denoted it as \textbf{MP}.
When LSTM outputs were used to calculate utility and penalty (without WC-CVaR), we denoted it as $\textbf{LP}$.
We compared our framework with several algorithms. 
A \textit{static algorithm} denoted as \textbf{ST} picked $m$ cells with the largest total demands in $D_1$, each assigning one static MU. This simulated the conventional EC scenario with static resources deployed to tackle fixed hotspots.
A \textit{greedy algorithm} denoted as \textbf{GD} used only the predictive mean of BLSTM, and picked $m$ cells with the largest average predictive means in $\omega_t$ to assign the MUs, one per cell.
We further implemented an out-of-box DRL 
algorithm, \emph{Synchronous Advantage Actor Critic}~\cite{a2c}, to dynamically learn the environment and make MU scheduling decisions, with two MLPs as actor and critic with $9155$ parameters in total.
Note that the \textbf{RL} took as input state the same predicted $\boldsymbol{\mu}$ and $\boldsymbol{\sigma}$ of all demand values within $\omega_t$ as our full algorithm \textbf{SP}, thus supposedly also possessing the ability to learn and utilize uncertainty information.
Denote the full algorithm with the greedy order of MU scheduling as \textbf{SPM}.

\subsubsection{Metrics}
We use the following metrics for evaluation.
\emph{Average profit} keeps track of the total profit each algorithm accumulated over the test dataset.
\emph{Number of violation cases} counts the number of cells where the observed excess demand (as defined in Eq.~\eqref{eq:excess}) is positive.
We also show trade-off between \emph{average utility} and \emph{average penalty} of our algorithms.
Finally, efficiency is measured in \emph{running time} per time step.

\subsection{Evaluation Results}

\begin{figure}[tb]
\centering
    \subfloat[Average Profit]{\includegraphics[width=0.47\textwidth, trim=0 10 0 0, clip]{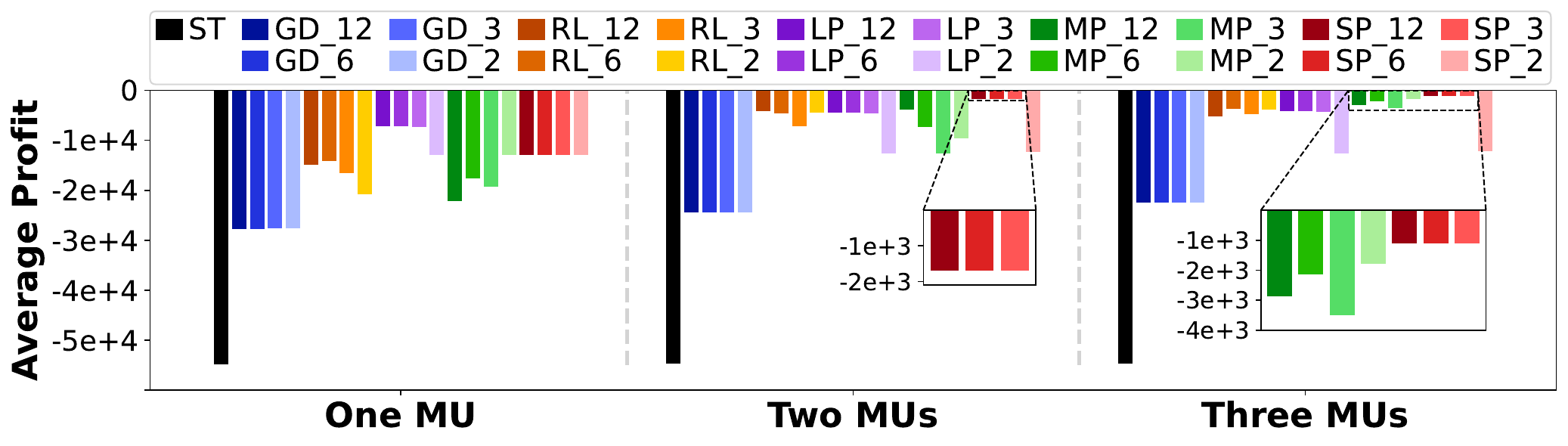}}
    \hfill
    \subfloat[Number of Violation Cases]{\includegraphics[width=0.47\textwidth, trim=0 10 0 0, clip]{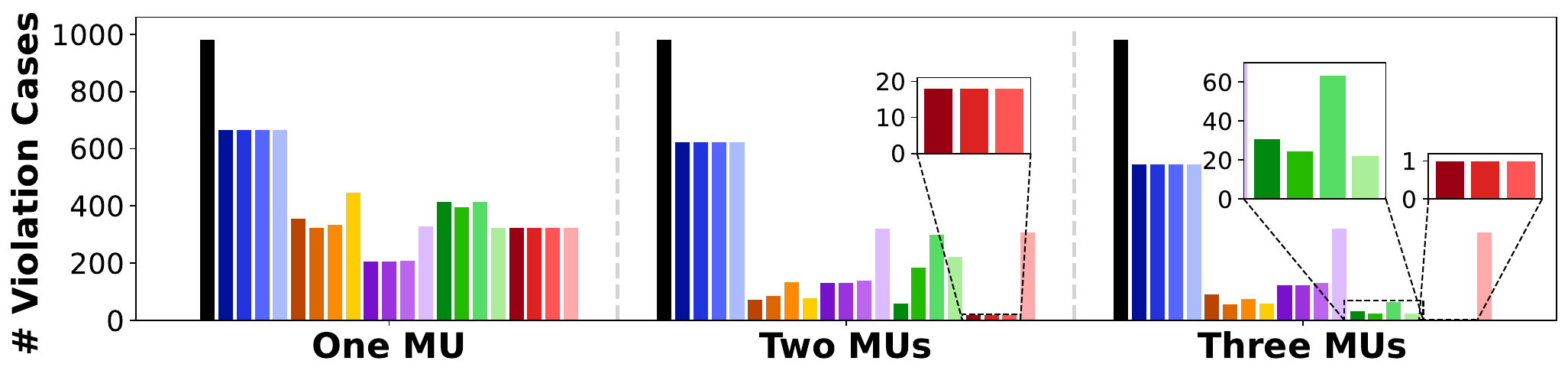}}
    \hfill
\caption{Average profit and number of total cases with excess demands (violations), with $m = 1, 2, 3$ and $L_{\mathsf{pred}} = 2, 3, 6, 12$ ($L_{\mathsf{pred}}$ marked after algorithm name)}
\label{fig:score_leftover}
\end{figure}

\subsubsection{Profit and excess demand comparison}
\noindent
In Fig.~\ref{fig:score_leftover}, we show the average profit and the number of violation cases with $m = 1, 2, 3$ MUs.
\textbf{ST} performed the worst which validates that static edge resources can hardly deal with demand surges.
\textbf{RL} only outperformed the baseline methods \textbf{ST} and \textbf{GD} in most cases.
This was due to the reward values \textbf{RL} tried to learn. 
Since we used a relatively large penalty constant for the SLA punishment, the reward with penalty could be much smaller than the one without.
This made the reward values almost a hard-to-learn sparse matrix. 
\textbf{LP} achieved the best with the largest profit and the smallest violation when there was only 1 MU in the system.
This was because when the resources were scarce and violation (penalty) was inevitable, LSTM could provide a more accurate point estimate of the demands which was beneficial for scheduling.
As the number of MUs added up to 3, \textbf{SP} performed the best while \textbf{MP} performed the second best of all algorithms.
\textbf{SP} specifically dominated when scheduling $3$ MUs, leading to notably higher profit and close to $0$ violation.
This was because {URANUS} considered data, model and distributional uncertainties, while using multi-agent planning to achieve close-to-optimal finite horizon scheduling.
We also observe no obvious advantage to ``look further into the future''. 
When $L_{\sf pred} = 2$, the planning horizon could not cover the maximum transit time between cells, leading to poor results. 
Beyond this, a small $L_{\mathsf{pred}}$ was sufficient for making satisfactory decisions, and increasing $L_{\mathsf{pred}}$ did not lead to obviously higher profit in general.
This justifies our \textbf{\emph{finite-horizon}} decision making.
\begin{figure}[t]
\centering
{\includegraphics[width=0.47\textwidth, trim=0 10 0 0, clip]{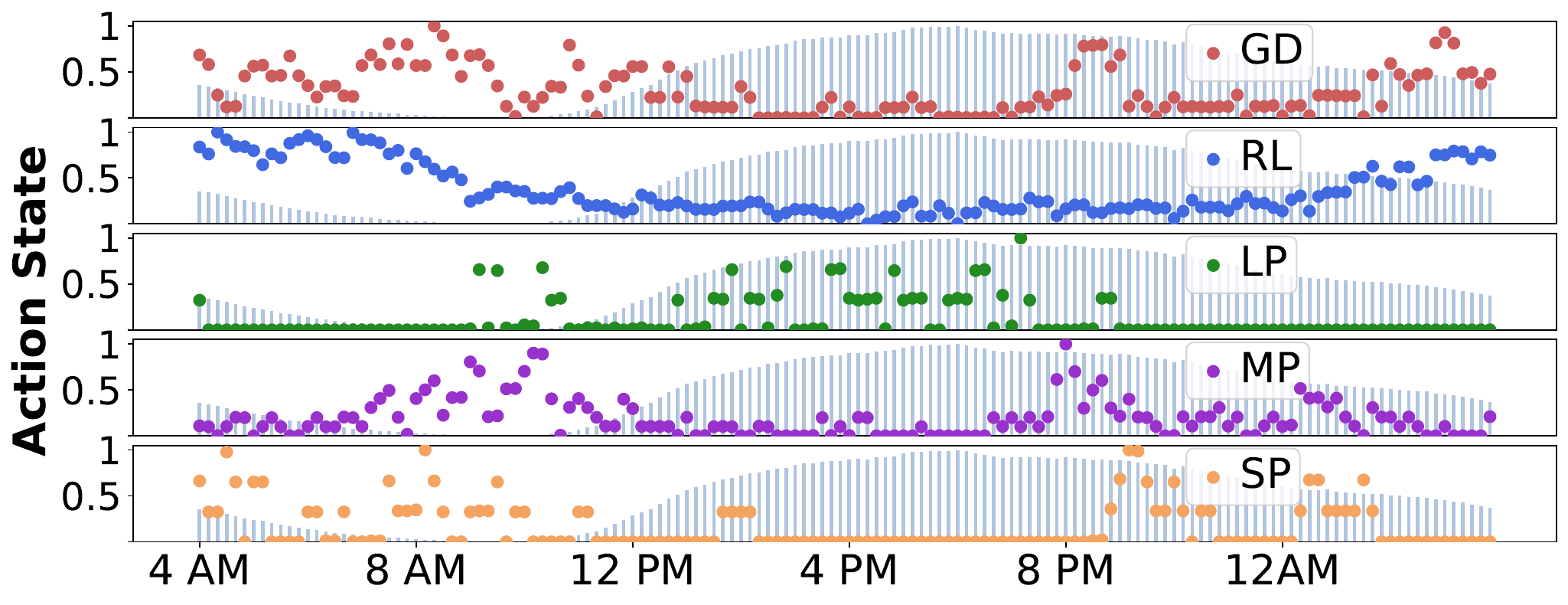}
}
\caption{MU action modes (colored dots) and demands (vertical bars) under different algorithms with $L_{\mathsf{pred}} = 12$}
\label{fig:action_algorithms}
\end{figure}

\begin{figure}[htb]
\centering
\subfloat[Average Profit]
{\includegraphics[width=0.246\textwidth, trim=0 0 0 0, clip]{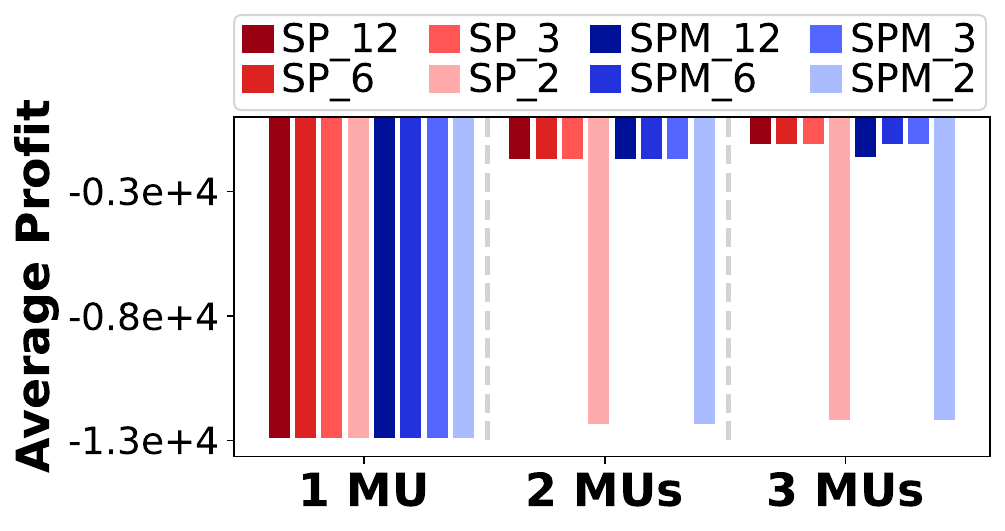}
}
\hfil
\subfloat[Number of Violation Cases]
{\includegraphics[width=0.223\textwidth, trim=0 0 0 0, clip]{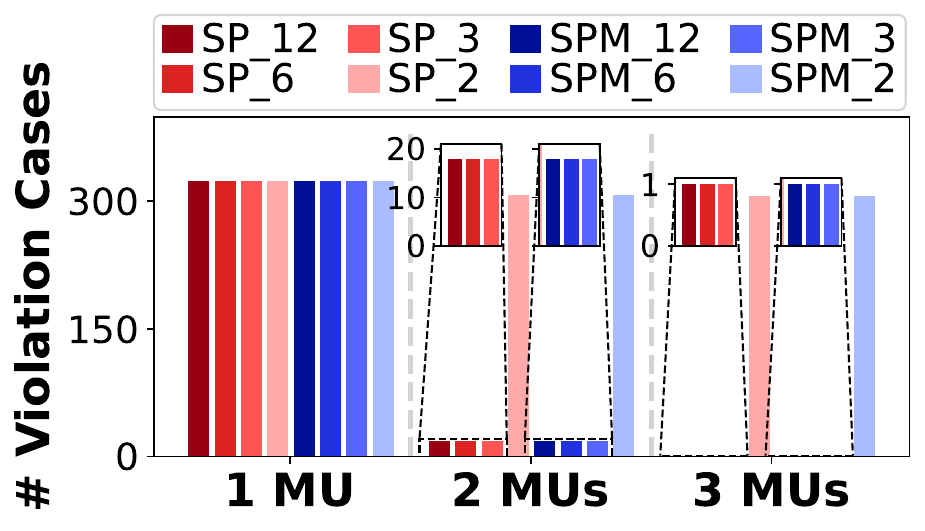}
}
\hfil
\caption{Average profit and number of total cases with excess
demands (violations) for \textbf{SP} and  \textbf{SPM}}
\label{fig:spm}
\end{figure}

In Fig.~\ref{fig:action_algorithms}, we show the correlation between \emph{actual demands} and \emph{actions} taken by each algorithm when scheduling 3 MUs to explain the superior performance of \textbf{SP} and \textbf{MP}.
Here, the \emph{action mode} represents the \emph{average ratio of MUs scheduled to be in transit in each time step within a day}, averaged over all days in $D_3$.
A lower value shows more \emph{stability} in keeping the MUs serving cells than being more in transit.
We can see that \textbf{SP} was overall the most stable among all, with a strong correlation between high demands and serving cells.
\textbf{MP} was the next most stabilized during peak hours.
Such stabilized serving indicates: a) accurate prediction of demands and/or rewards, b) close-to-optimal planning algorithm, and c) robustness against small demand fluctuations.
Compared to them, \textbf{RL} was less stabilized in that MUs might move even during peak hours, resulting in loss of profit.
\textbf{LP} surprisingly moved more during peak hours yet less during off hours.
After examination, this was due to lack of robustness against random fluctuations in predictions, causing constantly changing decisions when multiple cells might have roughly equal demands (and hence reward gains).
In Fig.~\ref{fig:spm}, we show the results for \textbf{SP} and \textbf{SPM}.
We observe that the scheduling order had negligible impact in \textbf{SP} versus \textbf{SPM}.
However, we indeed observe an increase in running time in \textbf{SPM} (not shown due to page limit).
This justifies using \textbf{SP} as the default mode of URANUS for comparing to the other algorithms above.

\subsubsection{Impact of penalty constant}
\begin{figure}[htb]
\centering
{\includegraphics[width=0.46\textwidth, trim=90 170 250 170, clip]{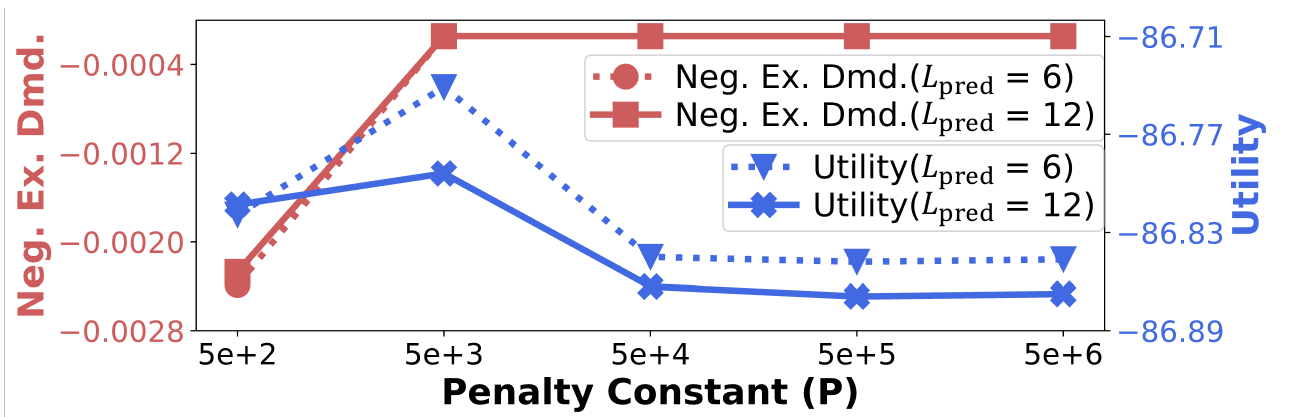}
}
\caption{Negative excess demand and utility with varying $P$}
\label{fig:penalty}
\end{figure}

\noindent
We further examined the impact of the penalty constant $P$ to show how algorithms adjust to the trade-off between utility and penalty.
We varied the penalty constant $P \! \in \! \{ 5\mathrm{e}{+6}, 5\mathrm{e}{+5}, 5\mathrm{e}{+4}, 5\mathrm{e}{+3}, 5\mathrm{e}{+2} \}$
and show the \emph{average profit} and the \emph{(negative) of the average total excess demand} as two sides of the trade-off for the \textbf{SP} algorithm in Fig.~\ref{fig:penalty}.
With increasing $P$, more focus had been put on reducing penalty, and hence the negative excess demand increased.
The utility values decreased after a certain threshold $5\mathrm{e}{+3}$, showing that the algorithm traded-off utility to reduce the increasingly large penalty.
The initial utility increase was due to increased robustness in utility estimation when taking the (relatively small) penalty into account.

\begin{figure}[b]
\centering
{\includegraphics[width=0.47\textwidth, trim=0 0 0 0, clip]{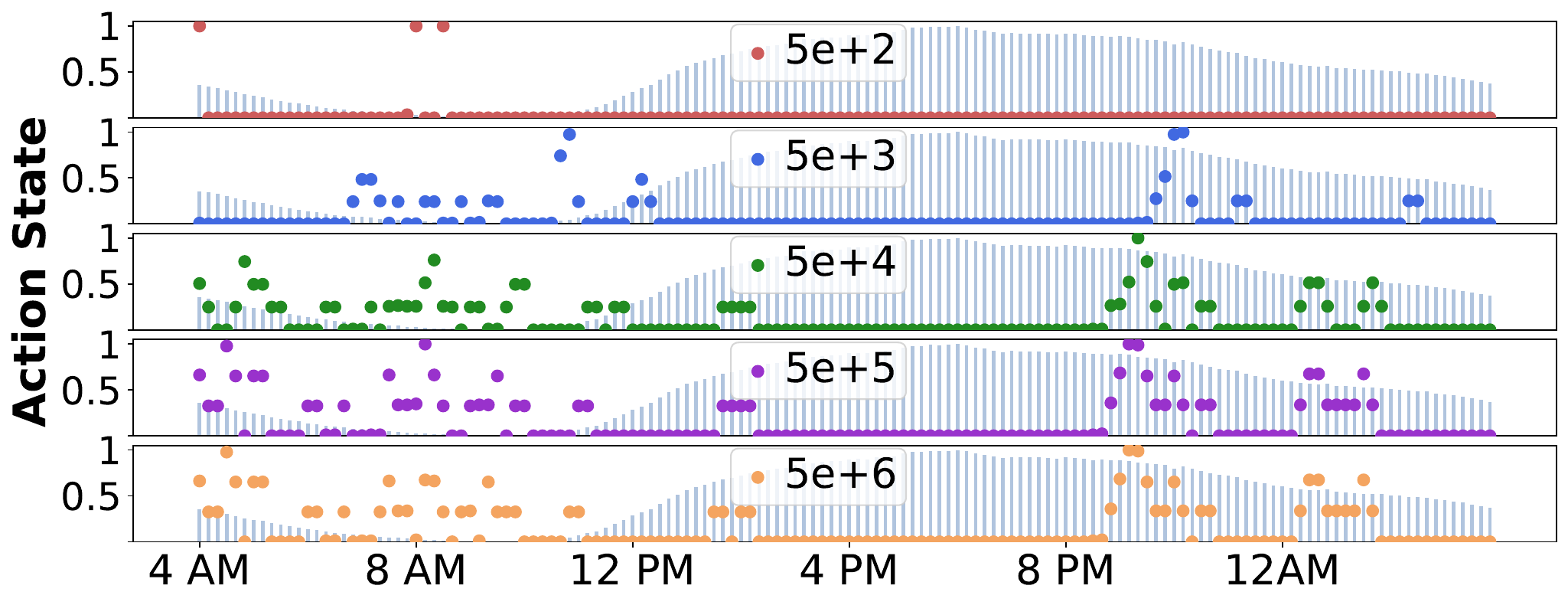}
}
\caption{MU actions versus demands with varying $P$ for \textbf{SP}}
\label{fig:action_sp}
\end{figure}

In Fig.~\ref{fig:action_sp}, we show the correlation between the \emph{actual demands} and \emph{actions} of \textbf{SP} under different $P$.
With a small $P$ such as $5\mathrm{e}{+2}$, the MUs chose to stay rather than move, due to the cost and wasted time during transit.
Movement increased when $P$ increased, since under-served excess demands might incur large penalties compared to the cost savings.
It stabilized beyond $P \!=\! 5\mathrm{e}{+4}$, when penalty started to dominate.
Another observation is that as demands went down and hence penalties were not incurred, the MUs kept moving to search for cells with highest utilities; when the demands were high, the MUs tended to stay and serve the nearby users.

\subsubsection{Efficiency}
\noindent
Table~\ref{tab:time} shows the time used for the resource scheduling task of all algorithms for different $L_{\sf pred}$.
Running time increased with the increase of $L_{\mathsf{pred}}$.
While achieving the best profit, \textbf{SP} bore somewhat heavy  running time for repetitively solving SDPs, though still within a reasonable decision time of several seconds given each scheduling period to be 10 minutes. 
Meanwhile, we found that major complexity comes from solving the SDP (accounting for over 99\% of the total time), which grows linearly with the number of locations and MUs, and can be further computed in parallel among all locations.
\textbf{MP}, with the second best results, significantly reduced running time, at the cost of slightly less performance and stability in planning as in Fig.~\ref{fig:score_leftover}..
Overall, combining learning-based prediction model with conventional robust optimization and combinatorial planning algorithm remains the most performant and stable approach for complex planning problems such as the one studied in this paper, despite the higher running time.

\begin{table}[htb]
\footnotesize
\centering
    \renewcommand\arraystretch{1.3}
    \setlength{\tabcolsep}{1.2mm}{
        \begin{tabular}{|c|c|c|c|c|}
        \hline
        \textbf{Algorithm}    & $L_{\mathsf{pred}}=2$ & $L_{\mathsf{pred}}=3$ & $L_{\mathsf{pred}}=6$ & $L_{\mathsf{pred}}=12$ \\ \hline
        \textbf{LP} & 0.7690 & 1.2961  & 2.8209  & 6.0417   \\ \hline
        \textbf{MP} & 6.5756  & 12.2388 & 29.8752  & 63.3239   \\ \hline
        \textbf{SP} & 743.5786  & 1431.0473  & 3431.2394  &  7427.3807  \\ \hline
        \end{tabular}}
        \caption{Running time (ms) used for decision making per time step}
    \label{tab:time}
\end{table}

\section{Conclusions}
\label{sec:conclusions}
\noindent
This paper proposed the concept of the Moving Edge, utilizing computing resources deployed on movable vehicles to provide quick response to edge demand surges.
A learning-based and end-to-end framework, {URANUS}, was proposed to schedule the moving computing units, maximizing edge provider's utility while minimizing its cost and potential penalty due to inadequate resources.
The core design principle of {URANUS} was to explicitly predict, model and optimize for the \emph{uncertainty} both intrinsically in the edge demands and due to modeling and prediction errors.
Based on observations from predicting real-world demand data, URANUS novelly applies and combines three techniques: a Bayesian neural network for uncertainty quantification, distributionally robust approximation to tackle distributional uncertainty, and finite-horizon planning to make optimized scheduling decisions.
Additional learning-based method was proposed to further improve efficiency.
With evaluation on real-world traffic dataset, we demonstrated that URANUS is uniquely suitable for realizing robust service-level agreements and avoiding violation penalty, and outperforms end-to-end reinforcement learning, uncertainty-agnostic planning, or heuristic baseline solutions.
Besides the studied scenario, the proposed uncertainty-aware framework hints new directions for combining learning-based methods, risk management, stochastic optimization, and optimal decision making under a unified framework.

{%%%%%%%%%%%%%%%%%%%%%%%%%%%%%%%%%%%%
%
%%%%%%%%%%%%%%%%%%%%%%%%%%%%%%%%%%%%
\bibliographystyle{myIEEEtranS}
\bibliography{IEEEabrv,ref}

}

\end{document}